\documentclass[letterpaper,singlespace]{aastex}
\usepackage{epsfig,lscape,natbib,emulateapj5}
\usepackage{color}
\def\lap{\lower.5ex\hbox{$\; \buildrel < \over \sim \;$}}
\def\gap{\lower.5ex\hbox{$\; \buildrel > \over \sim \;$}}

\def\ergcm2s{${\rm erg\ cm^{-2}\ s^{-1}}$}

\def\ergscm2s{${\rm erg\ cm^{-2}\  s^{-1}}$}

\def\cm-2{${\rm cm^{-2}}$}

\begin{document}

\title{Tracing the Metal-Poor M31 Stellar Halo with Blue Horizontal Branch Stars\\}

\author{Benjamin F. Williams\altaffilmark{1},
Julianne J. Dalcanton\altaffilmark{1},
Eric F. Bell\altaffilmark{2},
Karoline M. Gilbert\altaffilmark{3},
Puragra Guhathakurta\altaffilmark{4},
Claire Dorman\altaffilmark{4},
Tod R. Lauer\altaffilmark{5},
Anil C. Seth\altaffilmark{6},
Jason S. Kalirai\altaffilmark{3},
Philip Rosenfield\altaffilmark{7},
Leo Girardi\altaffilmark{8} 
}
\altaffiltext{1}{Department of Astronomy, Box 351580, University of Washington, Seattle, WA 98195; ben@astro.washington.edu; jd@astro.washington.edu; philrose@astro.washington.edu}
\altaffiltext{2}{Department of Astronomy, University of Michigan, 550
Church St., Ann Arbor MI 48109; ericbell@umich.edu}
\altaffiltext{3}{Space Telescope Science Institute; kgilbert@stsci.edu; jkalirai@stsci.edu}
\altaffiltext{4}{UC Santa Cruz; raja@uco.lick.org}
\altaffiltext{5}{NOAO; lauer@noao.edu}
\altaffiltext{6}{University of Utah; aseth@astro.utah.edu}
\altaffiltext{7}{Department of Physics and Astronomy G. Galilei, University of Padova, Vicolo dell'0sservatorio 3, I-35122 Padova, Italy; philip.rosenfield@unipd.it}
\altaffiltext{8}{Osservatorio Astronomico di Padova -- INAF, Vicolo dell'Osservatorio 5, I-35122 Padova, Italy; leo.girardi@oapd.inaf.it}

\begin{abstract}

We have analyzed new HST/ACS and HST/WFC3 imaging in F475W and F814W
of two previously-unobserved fields along the M31 minor axis to
confirm our previous constraints on the shape of M31's inner stellar
halo.  Both of these new datasets reach a depth of at least F814W$<$27
and clearly detect the blue horizontal branch (BHB) of the field as a
distinct feature of the color-magnitude diagram.  We measure the
density of BHB stars and the ratio of BHB to red giant branch stars in
each field using identical techniques to our previous work.  We find
excellent agreement with our previous measurement of a power-law for
the 2-D projected surface density with an index of 2.6$^{+0.3}_{-0.2}$
outside of 3 kpc, which flattens to $\alpha <$1.2 inside of 3 kpc.
Our findings confirm our previous suggestion that the field BHB stars
in M31 are part of the halo population.  However, the total halo
profile is now known to differ from this BHB profile, which suggests
that we have isolated the metal-poor component.  This component
appears to have an unbroken power-law profile from 3--150 kpc but
accounts for only about half of the total halo stellar mass.
Discrepancies between the BHB density profile and other measurements
of the inner halo are therefore likely due to the different profile of
the metal-rich halo component, which is not only steeper than the
profile of the metal-poor component, but also has a larger core
radius.  These profile differences also help to explain the large
ratio of BHB/RGB stars in our observations.

\end{abstract}


\section{Introduction}

Stellar halos of massive galaxies are likely formed in great part from
the disruption of dwarf galaxies.  The stellar content of these halos
thus offer quantitative insight into galaxy assembly
\citep[e.g.,][]{bullock2001,bullock2005,bell2008,Johnston2008,zolotov2009,cooper2010,Deason2013,Pillepich2014}. One
key observable property that is typically compared to simulations is
the radial stellar density profile.  These profiles are commonly
measured in nearby galaxies, such as the MW
\citep{juric2008,sesar2011}, M31
\citep{raja2005,courteau2011,Ibata2014}, N253 \citep{Bailin2011,
  Greggio2014} and M81 \citep{Barker2009}. However, such profiles have
proven difficult to measure over wide radial ranges.  In particular,
the profile shape of the inner parts ($\lap$10~kpc) of stellar halos
are difficult to probe.  This shape is important to determine as the
density profile at large radii can diverge if extrapolated to zero
radius.

Much work has gone into measuring the stellar density profile of
M31. M31 makes an excellent case study because it is very nearby, and
its halo has been resolved to 150~kpc
\citep[e.g.,][]{Ibata2001,kaliraihalo2006,gilbert2007,mcconnachie2009,gilbert2009streams,courteau2011,gilbert2012,williams2012,dorman2013,Ibata2014,gilbert2014}.
Recently, the Panchromatic Hubble Andromeda Treasury
\citep[PHAT][]{dalcanton2012} has started a new era in the study of M31 by
producing homogeneous HST imaging of about 1/3 of the star forming
disk.  One of the unexpected discoveries from this project was the
detection of field BHB stars within M31 bulge fields.  The surface
density of these BHB stars was found to follow the trend from the
outer halo, strongly suggesting that these stars belong to the halo
population and allow us to directly measure the inner halo stellar
density profile in M31 to previously-unattainable radii
\citep{williams2012}.

Our work with the PHAT data suggested that, if the BHB is indeed a
pure halo population, then the stellar halo has a surface density
profile with a 2-D power-law index of $\alpha=2.6^{+0.3}_{-0.2}$ that
flattens to $\alpha <$1.2 in the inner 3 kpc \citep{williams2012}, and
the BHB/RGB ratio of the halo component is relatively constant with
radius.  These findings about the BHB population of M31 produced
predictions of the BHB stars that would be observed in
previously-unobserved HST fields. If confirmed, these would show that
the power law continues to be steep to radii of $\sim$3~kpc and turns
over to $>{-}1.2$ at $<$3~kpc, the only measurement of a halo profile to
very small radii.

At the same time, there were a number of outstanding questions left
open by the BHB stellar halo measurement.  The interpretation of these
BHB stars traced the halo component hinged on the assertion of a
constant BHB/RGB ratio in the halo from 20~kpc (where it was measured
in halo only fields) to the inner parts.  Since the metallicity of the
stellar halo is known to vary with radius, this assertion left open
the possibility that inferences of the density profile of the inner
stellar halo based on BHB stars were incomplete.

Furthermore, recently there have been several relevant and
comprehensive studies of the M31 halo structure by \citet{dorman2013},
\citet{Ibata2014}, and \citet{gilbert2014}.  \citet{dorman2013}
decomposed the structure of M31 by simultaneously fitting surface
brightness, stellar luminosity function, and stellar velocity
data. They found a halo profile with a power-law index of 2.5$\pm$0.2,
very close to that measured using the BHB stars alone.  However, they
found a core radius, where the halo stellar profile flattens, to be
$\sim$10~kpc, apparently at odds with the BHB stars, which clearly
continue to follow the $\sim$2.5 power-law index well inside of
6~kpc. Confirming previous measurements of a metallicity gradient in
the M31 halo \citep{kalirai2006,koch2008,tanaka2010},
\citet{Ibata2014} found that the different metallicity components of
the M31 halo have significantly different profiles outside of 27~kpc,
so that measuring the metal-poor component, as we trace with the BHB,
does not necessarily tell the whole story.  \citet{gilbert2014}
quantified the metallicity gradient of the halo with
spectroscopically-confirmed members and included radii as small as
9~kpc.

In this paper, we probe BHB stars at intermediate radii, making it
possible to test for departures from a R$^{-2.6}$ power law suggested
by the outer fields and the inner fields under the assumption that the
BHB belongs to the halo component. We find that the BHB projected
number density matches a R$^{-2.6}$ profile to within the measurement
uncertainties at these radii as well, strengthening the interpretation
of the innermost BHB stars as probes of the metal-poor halo at small
radii.  This result combined with earlier measurements of the stellar
halo density at large radii \citep{gilbert2012,Ibata2014}, yield a
measurement of the metal-poor M31 stellar halo as an unbroken
power-law with a 2-D index of $\sim$2.6 over a very large radial range
(3--150 kpc). Furthermore, the results suggest that the difference in
the shapes of the metal-poor and metal-rich halo populations consists
of different core radii as well as different power-law indexes.  In
section 2 we describe our new data and provide a brief reminder of the
analysis technique.  Section 3 discusses the results and comparison
with predictions from previous work, and section 4 provides our
conclusions.  Throughout the paper, we assume a distance to M31 of 780
kpc \citep{stanek1998}, and a position angle of the major axis of
38$^{\circ}$ \citep{carignan2006}.

\section{Data}

\begin{figure}
\centerline{\epsfig{file=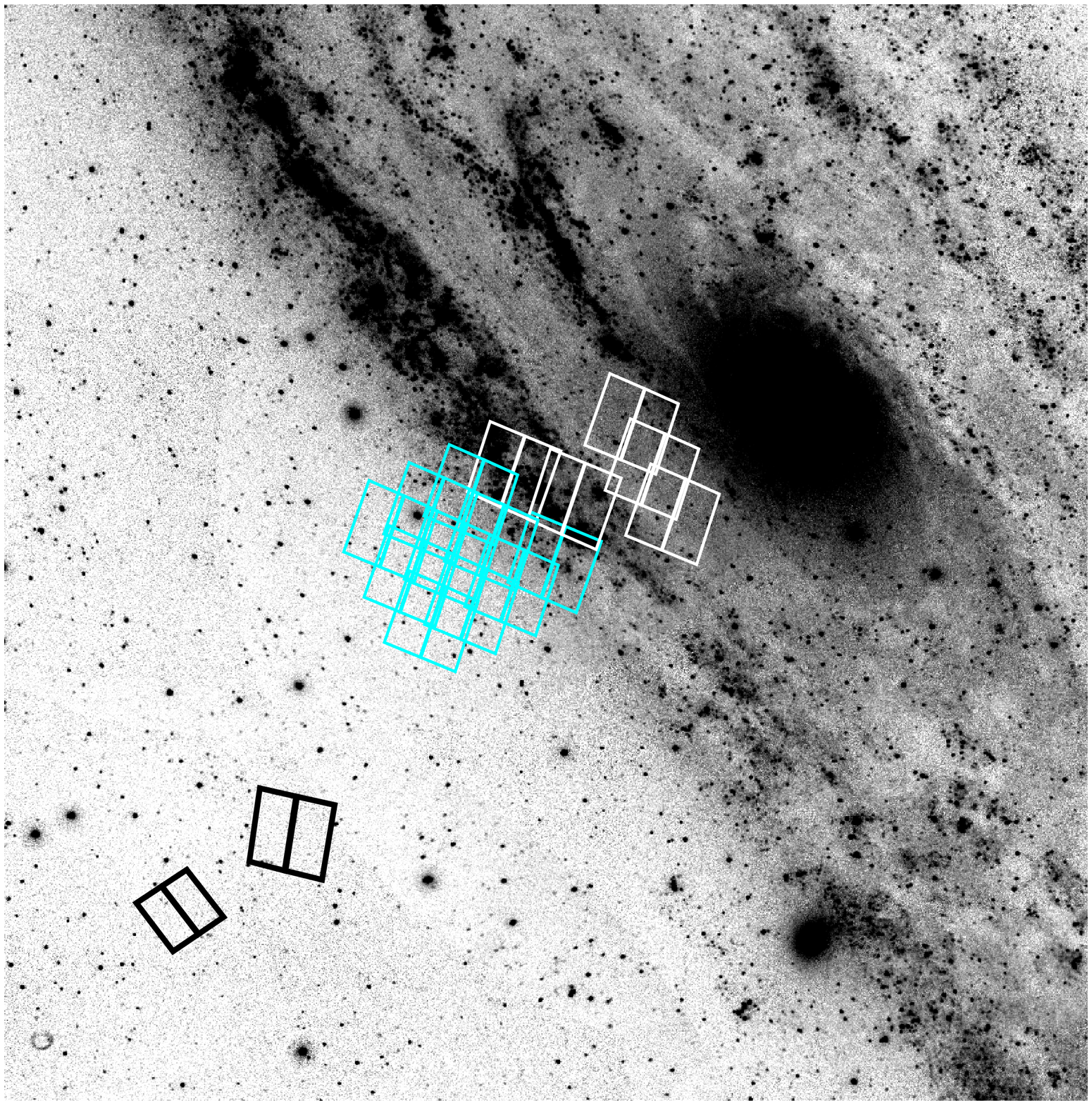,width=5.5in,angle=0}}
\caption{Black rectangles mark the location of our new ACS and UVIS
fields along the minor axis of M31.  Cyan rectangles mark the fields
from \citet{williams2012} that resulted in a BHB density measurement.
While rectangles mark the fields from \citet{williams2012} that
allowed BHB density upper limit measurements.}
\label{footprints}
\end{figure}

We observed the minor axis of M31 with HST for 2 orbits, using the ACS
as the primary camera and WFC3/UVIS as the parallel camera on 28
November 2013.  The telescope was oriented with PA\_V3=259 degrees,
which allowed both the ACS and UVIS fields to fall on the minor axis
simultaneously.  The footprints of the 2 fields are shown in
Figure~\ref{footprints}.  Each camera obtained one orbit of data in
F475W and one orbit in F814W.

Image calibration was preformed by the MAST reduction pipeline (OPUS
version 2013\_2) to produce the processed flt images on which we
performed our photometry.  The data were put through the same
photometry pipeline as that used for the PHAT photometry in
\citet{williams2012}, making the reduction identical to that of our
previous study.  Briefly, the flt images are put through multidrizzle
to identify cosmic rays; then they are multiplied by the pixel area
map and bad pixels and cosmic rays are masked to prepare the images
for photometry.  Finally the images are put through the DOLPHOT
\citep{dolphin2000} point spread function fitting routine to find and
measure magnitudes for all of the stars in the image.  DOLPHOT outputs
quality metrics such as signal to noise sharpness and crowding, which
are used to cull the results of artifacts and bad measurements.

\begin{figure}
\centerline{\epsfig{file=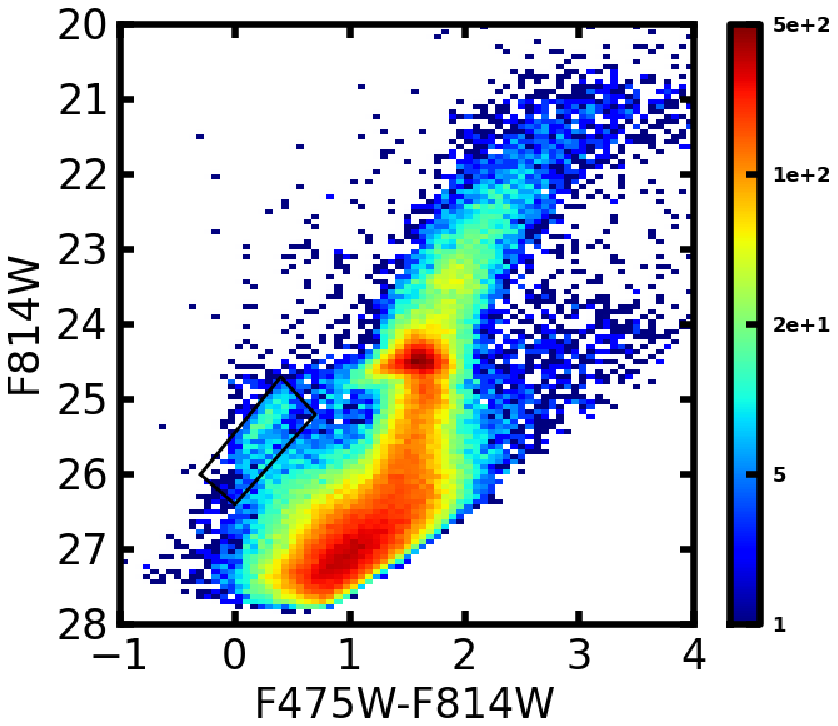,width=3.0in,angle=0}}
\centerline{\epsfig{file=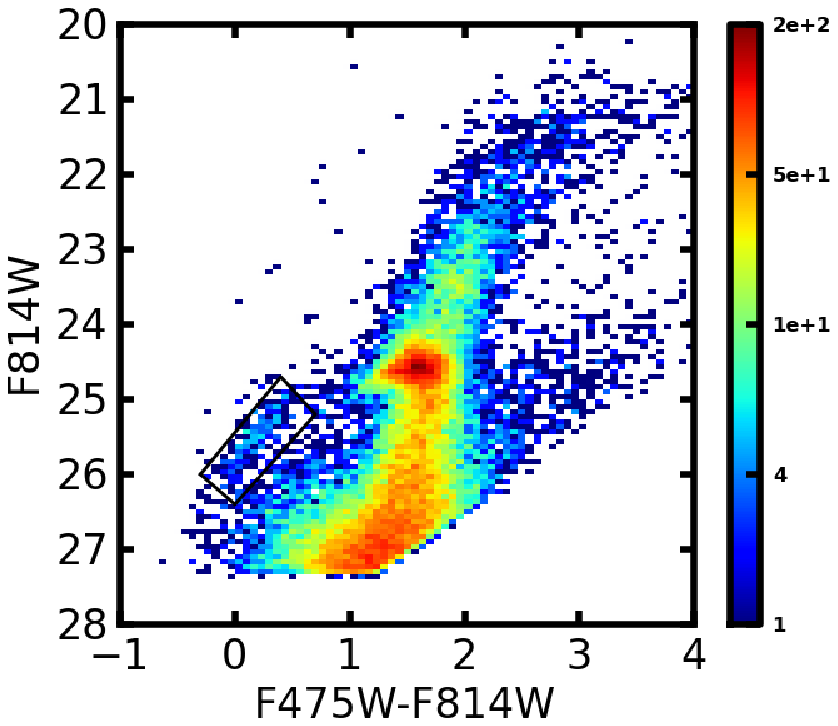,width=3.0in,angle=0}}
\caption{Color-magnitude diagrams (CMDs) from the ACS ({\it top}) and UVIS ({\it bottom}) imaging data. The BHB is outlined.  The UVIS field has significantly
fewer stars because it is smaller and farther out. }
\label{cmds}
\end{figure}

After this procedure 236561 and 61657 stars had been measured in the
ACS and UVIS fields, respectively, with signal-to-noise $>$4,
sharpness$^2{<}$0.2, and crowding $<$2.25 in both bands.  These
quality metrics are discussed in detail in \citet{dalcanton2012}.  The
resulting color-magnitude diagrams are shown in Figure~\ref{cmds},
where the smaller area of the UVIS camera as well as its slightly
larger galactocentric distance cause the lower number of stars in the
UVIS CMD.  The BHB is clearly visible in both CMDs; however, the
smaller number of BHB stars in the UVIS CMDs does result in somewhat
larger uncertainties, especially in the fraction of BHB to RGB stars.

\begin{figure}
\centerline{\epsfig{file=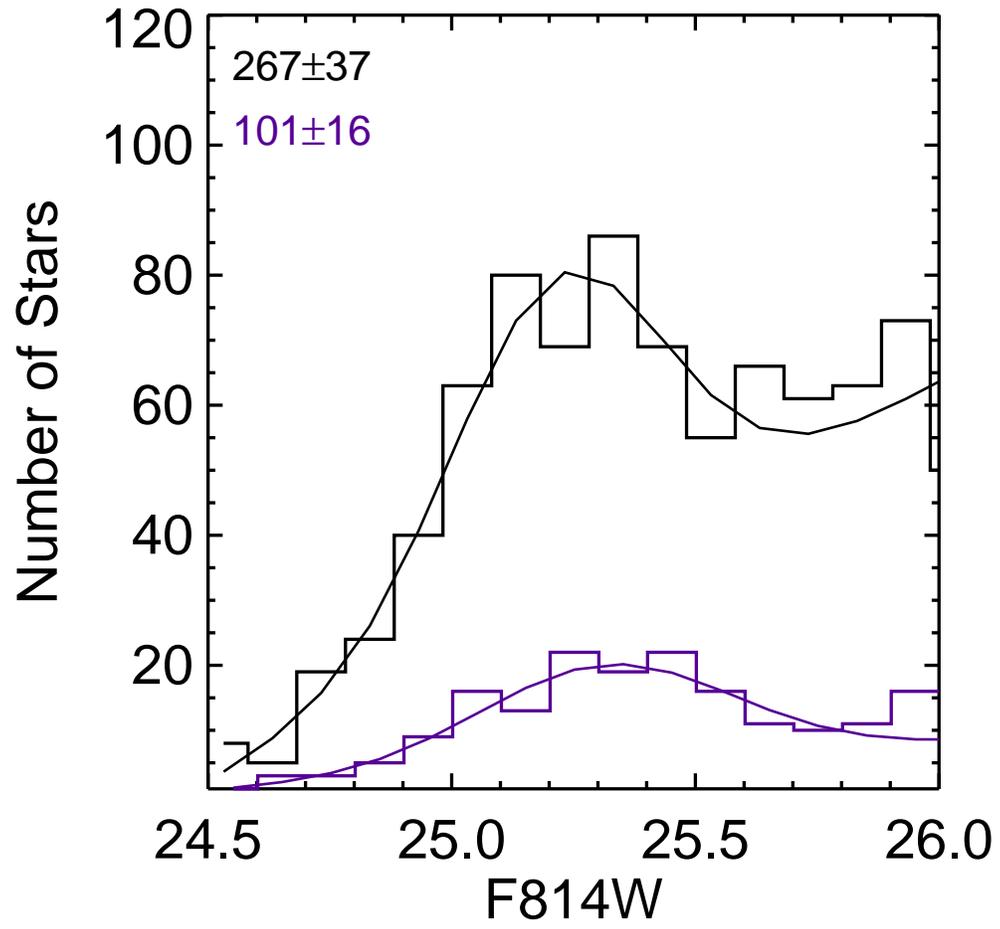,width=5.5in,angle=0}}
\caption{Fits to the luminosity function near the BHB in our two new HST fields.  A clear peak is detected, and the number of stars in the feature is given in the upper-left for each field.}
\label{fit}
\end{figure}

Once the photometry was complete, we determined the number of BHB and
RGB stars using the same method as \citet{williams2012}.  In short,
the number of BHB stars is measured by first fitting the F814W
luminosity function in the color range 0.1$<$F475W-F814W$<$0.5 with
the sum of a line and a Gaussian (see Figure~\ref{fit}).  Once fitted,
the integral of the line alone (non-BHB stars) is subtracted from the
integral of the fitted function to obtain the number of BHB stars.
The number of RGB stars was measured by counting the number of stars
in a box of CMD space where 22.0$<$F814W$<$22.5 and
1.5$<$F475W-F814W$<$3.5.

\section{Results and Discussion}

\begin{figure}
\centerline{\epsfig{file=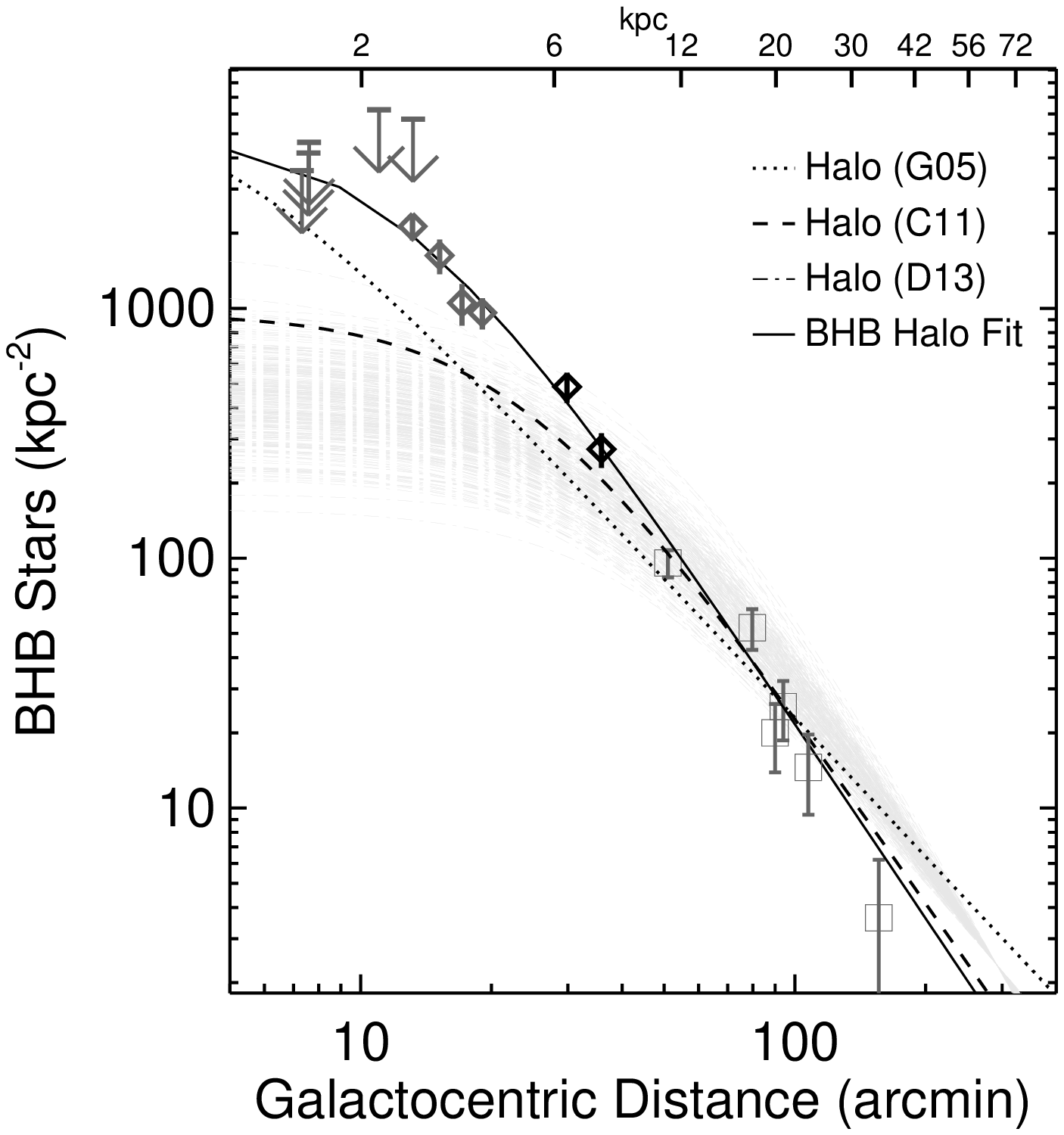,width=5.5in,angle=0}}
\caption{The number density of BHB stars measured for these fields
  (black data points at 30$'$ and 36$'$) overplotted on the radial
  distribution found in \citet[gray points;][]{williams2012}. Solid,
  dashed, and dotted lines mark the halo fits from
  \citet{williams2012}, from \citet{courteau2011}, and
  \citet{raja2005}, respectively.  Light gray dot-dashed lines mark a
  random draw of 256 MCMC results from \citet{dorman2013}.  The new
  data in the previously-unobserved region agrees with the model
  profile determined assuming that the BHB traces the low metallicity
  component of the stellar halo.}
\label{radial}
\end{figure}

\begin{figure}
\centerline{\epsfig{file=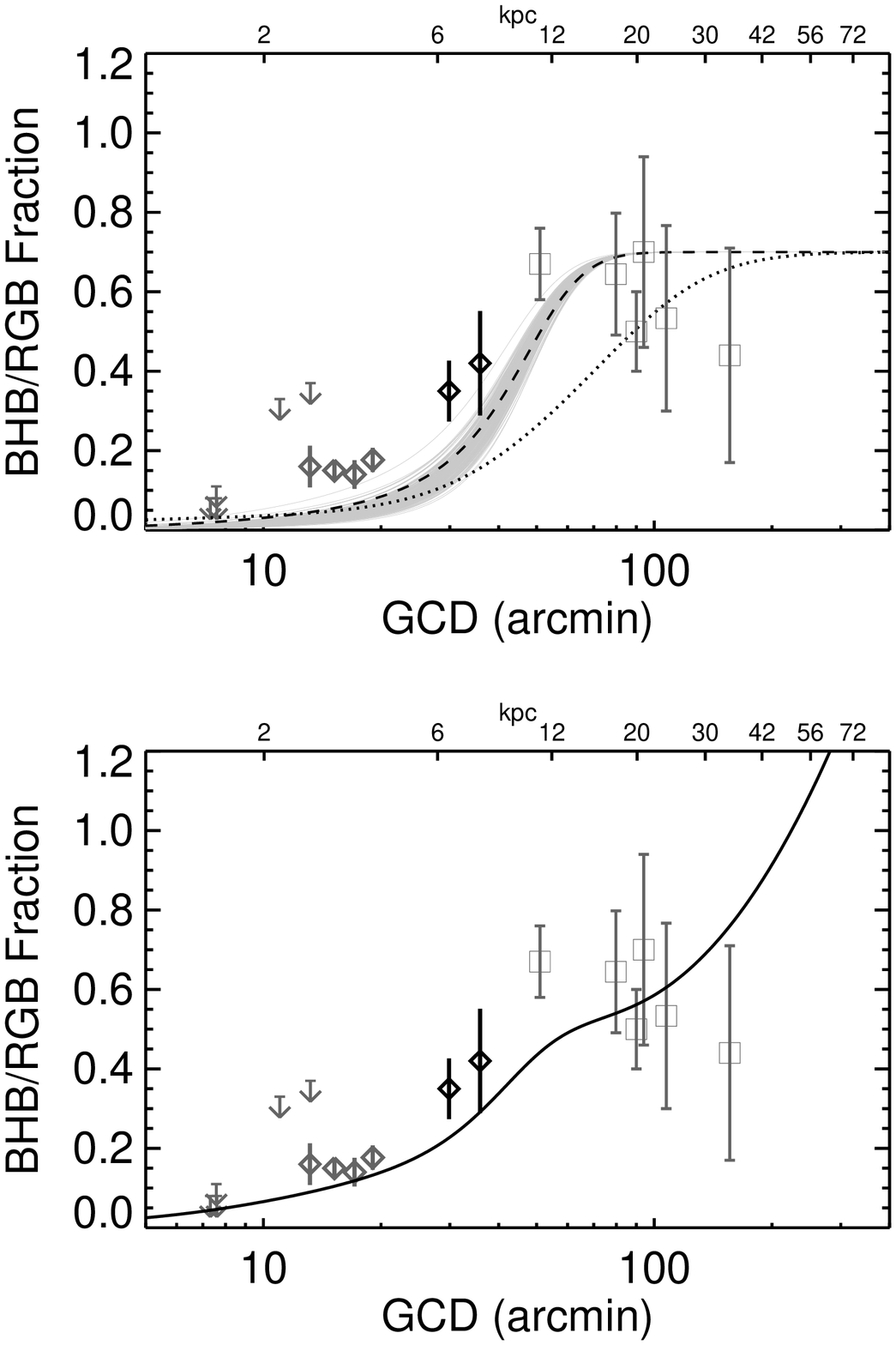,width=3.5in,angle=0}}
\caption{{\footnotesize The BHB/RGB ratio as a function of radius.
    Gray points are from \citet{williams2012}.  Black points are from
    the new observations at 30$'$ and 36$'$.  {\it Top:} Dashed, and
    dotted lines assume a constant BHB/RGB ratio of the pure halo of
    0.7 \citep{williams2012} and the halo fractions as a function of
    radius from \citet{courteau2011}, and \citet{raja2005},
    respectively. Gray lines also assume a constant BHB/RGB ratio of
    the pure halo of 0.7 but the halo fractions come from 256 trials
    from the \citet{dorman2013} analysis.  The strong transition
    between dominant components was predicted to occur in the
    previously-unobserved region from 20$'$ to 60$'$, and indeed, the
    new observations (at 30$'$ and 36$'$) follow the predicted sharp
    transition. {\it Bottom:} To match the sharp transition in BHB/RGB
    ratio, as well as the inner data points, requires an increase in
    the fraction of metal-poor halo stars inside of 10~kpc, as shown
    with the black solid line.  This line is the result of
    extrapolating of the metal-rich halo profile of \citet{Ibata2014}
    inward, adopting the stellar halo core radius measured by
    \citet{dorman2013} for the metal-rich component core radius, and
    adopting the BHB core radius and power-law index for the
    metal-poor halo component. Even this two-component halo model does
    not fully match the data; however, it is a significant improvement
    over any single-metallicity option, assuming previously-measured
    decomposition parameters.}}
\label{fraction}
\end{figure}

We have placed the measured BHB star densities and BHB/RGB fractions
from our two new fields on the same plots as \citet{williams2012} in
Figures~\ref{radial} and \ref{fraction}.  The lines on the plots are
not fits with the new data, but the same lines used in the previous
analysis.  These lines correspond to the halo profiles measured by
\citet{williams2012}, \citet{courteau2011}, \citet{raja2005}, and
\citet{dorman2013}.  The new observations are consistent with the fits
to the data from \citet{williams2012} while probing a
previously-unobserved radius.  The fact that the density falloff
mimics the expected drop from the inner radii to the outer radii shows
that the BHB component of these fields are not strongly affected by
substructure.

More importantly, the earlier data showed a steep transition at these
radii from low BHB/RGB ratio at the inner radii, where the RGB was
highly contaminated by disk and bulge stars, to a relatively constant
ratio of 0.7.  The previous measurements resulted in a prediction of a
smooth increase in BHB/RGB ratio at these radii to the constant 0.7
value. Indeed, the new observations confirm that the ratio follows the
predicted smooth transition; however, new studies have since been
published on the metallicity properties of the halo with radius.  In
light of these studies, the interpretation of this transition is more
complicated.

Given that the BHB stars are tracing the metal-poor stellar halo, it
is difficult to reconcile the fact that their density distribution
clearly follows a continuous power-law to galactocentric distances
inside of 6 kpc while sophisticated decompositions based on a much
larger amount of data suggest that the stellar halo density
distribution flattens outside of 8 kpc
\citep{courteau2011,dorman2013}.

Since our technique uses BHB stars, it necessarily only probes the
metal-poor component of the stellar halo, suggesting that this
component has a smaller core than the total stellar halo.  Over the
past decade, M31's halo has been shown to have a metallicity gradient
\citep{kalirai2006,koch2008,tanaka2010}.  Recently, \citet{Ibata2014}
found that the ``unmasked\footnote{measured without applying their
  masks to the substructure}'' projected stellar density profile of
the outer halo of M31 is a strong function of metallicity, with the
more metal-rich component ([Fe/H]${>}-$1.1) having a steep fall-off
($\alpha{\sim}$3.7) at large radii.  Their lower metallicity profiles
([Fe/H]${<}-$1.1) have similar slopes to the BHB stars
(2.3${<}\alpha{<}$2.7).  Thus it is possible that the discrepancy
between the BHB results and the kinematic and surface brightness
results for the inner regions is attributable to the metal-rich and
metal-poor components of the stellar halo having different profiles.
This possibility affects the use of the BHB in understanding both the
mass and profile of the M31 stellar halo.

Below we explore the possibility that the explanation for the
discrepancy between the multiple profile measurements is metallicity.
Herein, we simplify the problem by breaking the halo into 2
components: one metal-rich and one metal-poor.  M31's halo is likely
more complex than this simple two-component model--- the distribution
of metallicities in present in the population with radius is a mixture
of stars with a range of metallicities.  However, the BHB are likely
to mainly represent the metal-poor populations in the halo.  As shown
below, the large core radius measured by \citet{dorman2013} and the
small core radius measured using the BHB stars can be somewhat
reconciled if (1) the observed differences between the profiles of the
metal-rich and metal-poor stars persist into the inner regions of the
halo, and (2) the large core radius measured by Dorman et al. is
comprised primarily of metal-rich halo stars.

\subsection{The Stellar Halo Profile}

Recent work confirms that the power-law profile of the metal-poor halo
continues all the way out to at least 150~kpc
\citep{gilbert2012,Ibata2014}.  Thus, our mass estimate is valid for
the metal-poor ([Fe/H]$<$-0.7) component of the halo.  In particular,
\citet{Ibata2014} find that $\sim$50\% of the halo stellar mass from
27--150 kpc to have [Fe/H]${<}-$0.6 (see their Table 4).  The fraction
is comparable to the \citealp{brown2008} SFH result.  Clearly, a large
amount of the halo stellar mass is metal-rich, which could change the
characteristics of the core radius for the total halo.  Thus, while
the metal-poor halo appears to have a consistent power-law index of
${\sim}{-}2.5$ from 3 to 150 kpc, the total halo is more complex, with
multiple power-law indices, a larger core, and a larger mass.

\begin{figure}
\epsfig{file=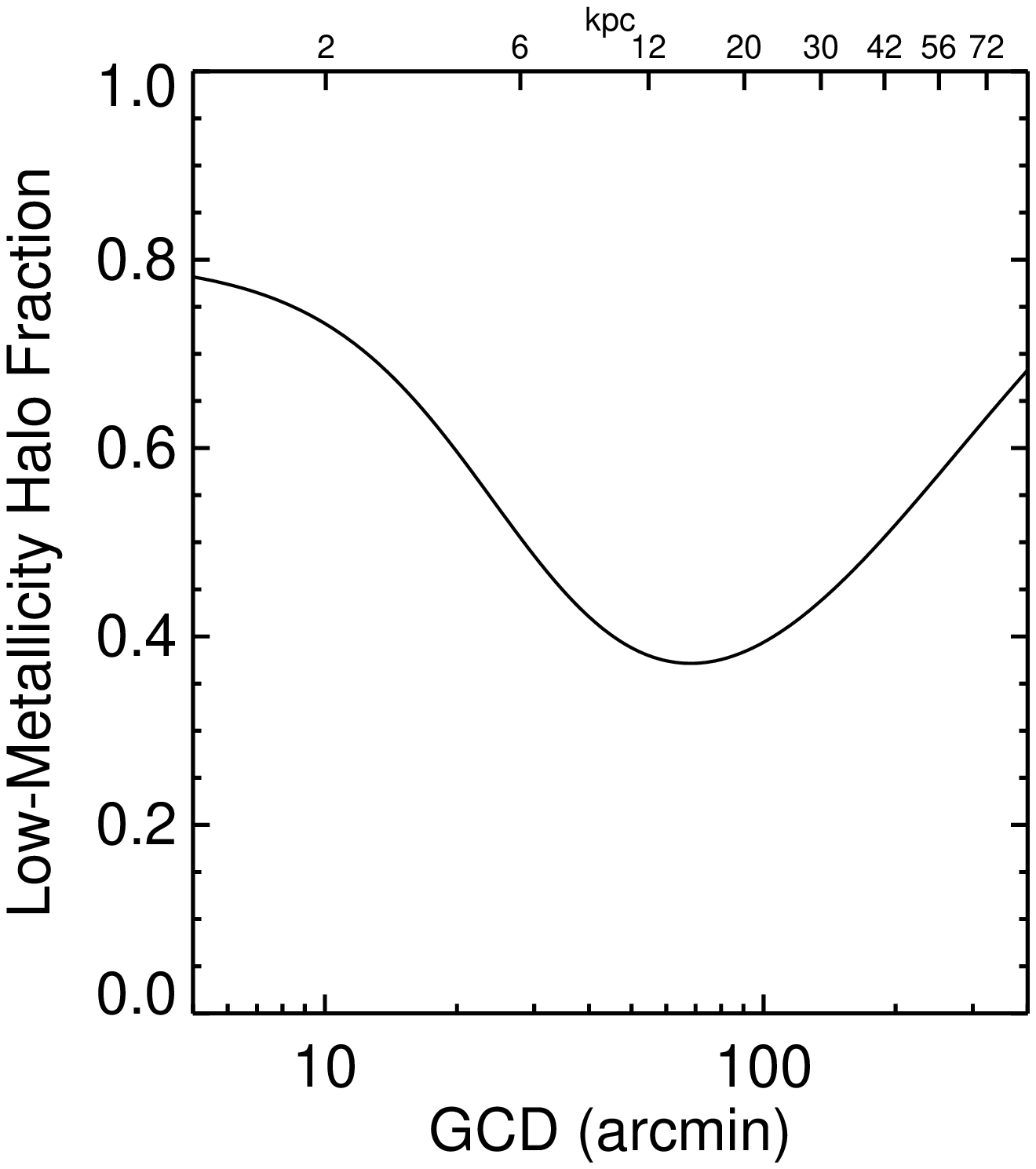,width=3.3in,angle=0}
\epsfig{file=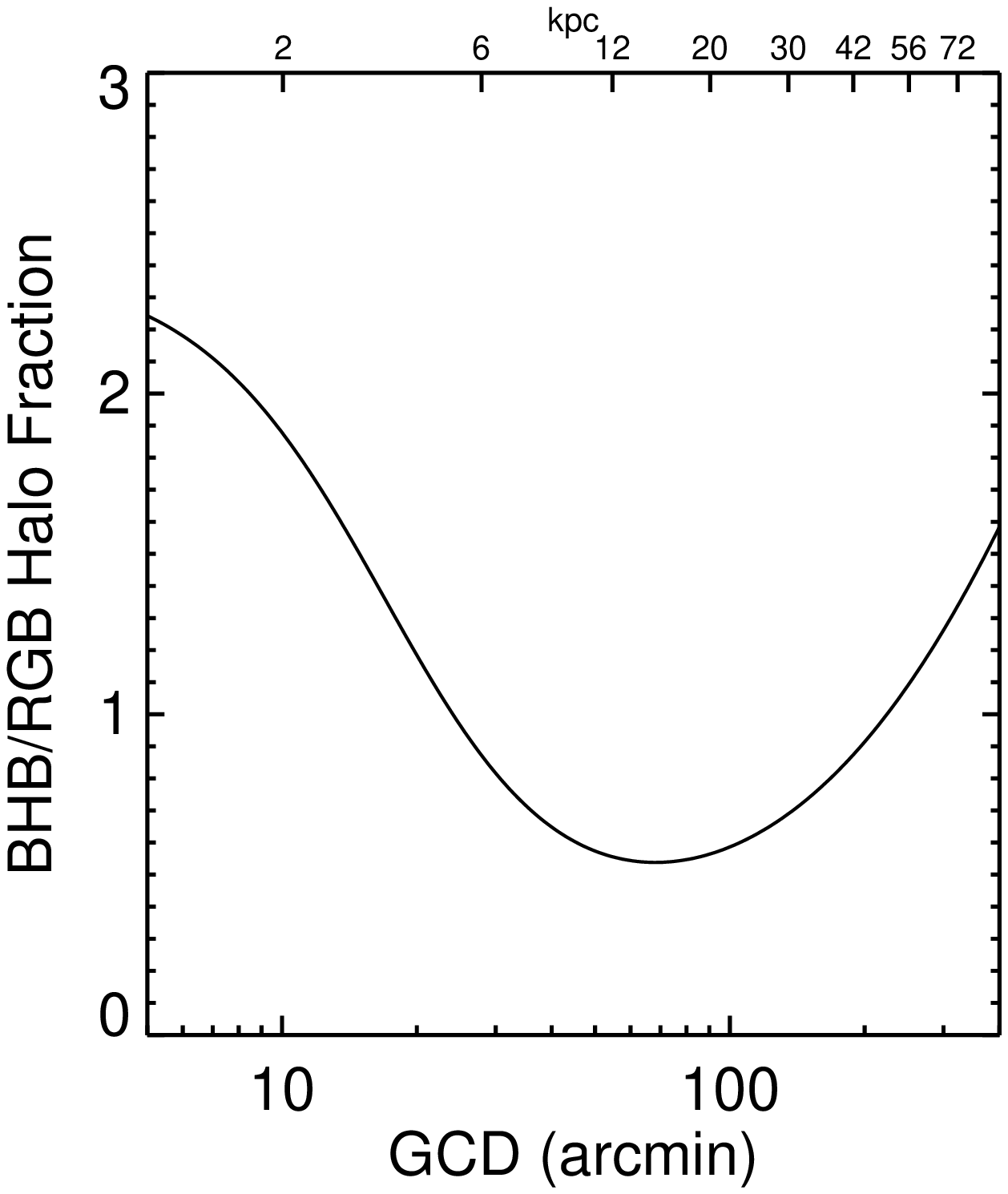,width=3.3in,angle=0}
\caption{{\it Left:} The fraction of metal-poor ([Fe/H]$<$-0.7) halo stars as a
  function of galactocentric distance predicted by extrapolating the
  \citet{Ibata2014} metal-rich halo profile, adopting the
  \citet{dorman2013} core radius for the metal-rich component, and
  adopting the BHB profile for the metal-poor component.  Because of
  the smaller core radius of the metal-poor component, the metal-poor
  fraction recovers inside of the metal-rich core radius.  {\it Right:} The predicted BHB/RGB ratio of the halo alone as a function
  of galactocentric distance, assuming the metal-poor component has a
  ratio of 5, comparable to metal-poor globular clusters
  \citep{williams2012}.  The high value in the inner galaxy is again
  due to the smaller core radius of the metal-poor component.}
\label{fracs}
\end{figure}

If the large core radius of the total stellar halo is the result of
the high-metallicity component, then one might expect the BHB/RGB
ratio of the halo component to increase at smaller radii as the ratio
of metal-poor stars to total stars increases.  In
Figure~\ref{fracs} we plot the expected ratio of metal-poor stars
to total stars with galactocentric distance.  We assume that the
metal-poor component of the halo has a profile with a core radius of
3.5~kpc and a 2D power-law index of 2.7, and the metal-rich component
has a core radius of 10.6~kpc and a 2D power-law index of 3.9.  The
result is that the metal-poor fraction has a minimum at 10-20~kpc, and
a steep increase toward the galaxy center.  These ratios are similar
to those found by \citet{gilbert2014}, which were $\sim$0.2--0.7
between 9 and 30 kpc, respectively.  We are not able to match the
spectroscopic ratios exactly because the difference in power-law index
between the two components does not result in the fraction changing
that steeply.  We note that the spectroscopic sample in this radial
range contains a significant amount of substructure
\citep{gilbert2007}, which may be causing the rapid change in the
ratio of low-metallicity stars to total stars in those measurements.
Therefore, we set the relative normalization between the two halo
components so that in this radial range we roughly agree with the
spectroscopic ratio of low metallicity stars to total stars.

If we assume a BHB/RGB fraction of $\sim$5, similar to low-metallicity
globular clusters \citep{williams2012}, for the metal-poor component,
and a fraction of 0 for the metal-rich component, then we can convert
the metal-poor fraction into an expectation for the BHB/RGB ratio of
the halo as a whole with galactocentric distance.  This relation is
shown in Figure~\ref{fracs}, and roughly follows the metallicity
fraction.

Using these fractions along with the disk and bulge profiles from
\citet{dorman2013}, we calculate the expected BHB/RGB ratios for the
total population, which are plotted in Figure~\ref{fraction}.  This
plot assumes that the metal-poor halo is the only component that has a
significant BHB, and all components contribute to the RGB.  The
dashed, dotted, and gray lines assume a constant BHB/RGB halo fraction
of 0.7 and the profiles of \citet{courteau2011,raja2005}, and
\citet{dorman2013}, respectively.  The solid line assumes the BHB/RGB
fraction of the halo component follows the relation shown in
Figure~\ref{fracs}, and the total profile follows that of the median
fit from \citet{dorman2013}.  Although still not a perfect match to
the data, this relatively simple attempt to account for the changing
metallicity of the halo moves the model prediction considerably closer
to the data in the inner regions.  Therefore, one way to reconcile the
measurements is through an increasing metal poor fraction inside of
the \citet{dorman2013} core radius.

Therefore, one explanation for the difference between the core radii
of the BHB profile and the total halo profile could be that the
metal-poor halo has a significantly smaller core than the metal-rich
halo.  Such a possibility may be testable in simulations, as many of
the oldest, most metal-poor stars in the halo may actually form early
and largely in the center of the dark matter halo.  At the same time,
if much of the metal-rich halo is formed later through accretion
events, it could perhaps be hotter, resulting in a larger core.  Our
result can be compared to future simulations in this way.

\subsection{The Stellar Halo Mass}

If there is a significant amount of mass in the more metal-rich
component of the halo which is not probed by the BHB stars, our
previously-determined stellar halo mass of
2.1$^{+1.7}_{-0.4}{\times}10^9$ M$_{\odot}$ was not properly
calibrated, as the profile was normalized to the total stellar surface
brightness at 21~kpc ($\mu_V{\sim}29$) from the photometry of
\citet{brown2008}.  Based on observations of globular clusters as well
as stellar evolution theory, the BHB traces the metal-poor component
of the stellar population.  According to the star formation history
(SFH) in \citet{brown2008}, about 50\% of the stellar population has
[Fe/H]$\lap$-0.7.  If we assume a $V$-band stellar mass-to-light ratio
of of 2 for the metal-poor component \citep{bruzual2003,chabrier2003},
and normalize by the fraction of metal-poor stars (0.5), our
metal-poor halo stellar mass measurement is
1.5$^{+1.1}_{-0.3}{\times}10^9$ M$_{\odot}$. Since this only includes
the metal-poor component, it is a lower-limit on the total stellar
mass of the halo.

We can move beyond a lower limit if we assume the metal-rich component
has a core radius of 10~kpc, as measured by \citet{dorman2013} and the
steep 2-D power-law slope of -3.72 measured by \citet{Ibata2014} for
the unmasked metal-rich component.  If we assume a $V$-band stellar
mass-to-light ratio of 3 for the metal-rich component
\citep{bruzual2003,chabrier2003} and normalize the metal-rich
component to comprise the other half of the surface brightness at
21~kpc, then the mass of this component is $\sim$1.4${\times}10^9$
M$_{\odot}$.  Adding this to the mass of the metal-poor component then
yields a total of $\sim$3${\times}10^9$ M$_{\odot}$.  This estimate is
similar to our earlier estimate which did not properly account for the
metal-rich component.  This mass is about a factor of 2 lower than the
estimates of \citet{Ibata2014}.  Their mass estimates do not assume a
mass-to-light ratio, but instead fit the observed stars to correct to
the full underlying populations.\footnote{The luminosities are
  measurements within their selection function.} However, if we apply
our assumed mass-to-light ratios of 2 and 3 directly to their measured
$V$-band luminosities for the metal-rich and metal-poor components of
the outer halo, the masses of these components are 2$\times$10$^8$ and
4$\times$10$^8$ M$_{\odot}$, respectively.  Summing these values and
extrapolating inward with an assumed core radius of 10~kpc yields an
estimated total halo mass of $\sim$2$\times$10$^9$.  Thus, when the
same assumptions are applied to both measurements, our halo stellar
mass estimates are similar. However, ours is somewhat greater, largely
due to the smaller core radius of the metal-poor component.

\section{Conclusions}

We have tested the predictions for the distribution of BHB stars from
30--40 arcmin (6--10 kpc) away from the center of M31 along the minor
axis.  The latest observations confirm the predictions based on a
model where the BHB stars trace the metal-poor M31 stellar halo, and
the metal-poor stellar halo has a structure that follows a power law
with 2-D index $\alpha=2.6^{+0.3}_{-0.2}$ inward to 3~kpc, then
becomes significantly flatter near the galaxy center.  This
core-radius is significantly smaller than that measured for the total
stellar halo.  The BHB/RGB ratio we measure inside of 10~kpc suggests
that the metal-poor halo component may become increasingly important
inside of the total halo core radius of 10~kpc measured by
\citet{dorman2013}.  Therefore, not only does the slope of the stellar
halo profile change with metallicity, the core radius appears to
change with metallicity as well.

Combining this result with recent measurements of the total halo and
metal-rich halo, suggests that the stellar halo mass calculated in
\citep{williams2012} was likely underestimated because it did not
properly account for the fact that the BHB stars were only probing the
metal-poor component of the stellar halo.  A more realistic
calculation uses the metal-poor component, assuming a mass-to-light
ratio of 2, to set a lower-limit on the stellar halo mass of
1.5$^{+1.1}_{-0.3}{\times}10^9$ M$_{\odot}$. If we estimate the mass
of the metal-rich component using the \citet{Ibata2014} profile,
assuming a metal-rich mass-to-light ratio of 3, and normalizing to the
surface brightness at 21 kpc, then we can add this to the metal-poor
component estimate to derive a total stellar mass of the M31 halo is
$\sim$3${\times}10^9$ M$_{\odot}$.  The significant differences
between the metal-rich and metal-poor halo characteristics in M31
should be reflected in galaxy formation and evolution simulations.

Support for this work was provided by NASA through grant GO-12997 from
the Space Telescope Science Institute, which is operated by the
Association of Universities for Research in Astronomy, Inc., for NASA,
under contract NAS 5-26555.


\clearpage 

\end{document}